\documentclass[aps,pre,preprint,groupedaddress]{revtex4-1}
\newcounter{defcounter}
\usepackage{amsmath}

\newenvironment{myequation}{%
\addtocounter{equation}{-1}
\refstepcounter{defcounter}

\begin{equation}}
{\end{equation}}

\newcounter{matriz}

\usepackage[normalem]{ulem}
\usepackage{graphicx}
\usepackage{color}

\begin{document}

% Use the \preprint command to place your local institutional report
% number in the upper righthand corner of the title page in preprint mode.
% Multiple \preprint commands are allowed.
% Use the 'preprintnumbers' class option to override journal defaults
% to display numbers if necessary
%\preprint{}

%Title of paper
\title{Cooperation in two-dimensional mixed-games}

% repeat the \author .. \affiliation  etc. as needed
% \email, \thanks, \homepage, \altaffiliation all apply to the current
% author. Explanatory text should go in the []'s, actual e-mail
% address or url should go in the {}'s for \email and \homepage.
% Please use the appropriate macro foreach each type of information

% \affiliation command applies to all authors since the last
% \affiliation command. The \affiliation command should follow the
% other information
% \affiliation can be followed by \email, \homepage, \thanks as well.
\author{Marco A. Amaral$^{ 1\ast}$, Lucas Wardil$^{2}$ and
Jafferson K. L. da Silva$^{1}$}
%\email[]{Your e-mail address}
%\homepage[]{Your web page}
%\thanks{}
%\altaffiliation{}
\affiliation{
$^{1}$Departamento de F\'\i sica, Universidade Federal de Minas Gerais, \\
Caixa Postal 702, CEP 30161-970, Belo Horizonte - MG, Brazil\\
\\
$^{2}$Department of Mathematics, University of British Columbia,\\
1984 Mathematical Road, Vancouver B.C., Canada V6T1Z2\\
}
\email{marcoantonio.amaral@gmail.com}
%Collaboration name if desired (requires use of superscriptaddress
%option in \documentclass). \noaffiliation is required (may also be
%used with the \author command).
%\collaboration can be followed by \email, \homepage, \thanks as well.
%\collaboration{}
%\noaffiliation
\setlength{\parindent}{2cm}
\date{\today}

\begin{abstract}
Evolutionary game theory is a common framework to study the evolution of cooperation, where it is usually assumed that the same game is played in all interactions. Here, we investigate a model where the game that is  played by two individuals is uniformly drawn from a sample of two different games. Using the master equation approach we show that the random mixture of two games is equivalent to  play the average game when (i) the strategies are statistically independent of the game distribution and (ii) the transition rates are linear functions of the payoffs. We also use Monte-Carlo simulations in a two dimensional lattice and mean-field techniques to investigate the scenario when the two above conditions do not hold. We find that even outside of such conditions, several quantities characterizing the mixed-games are still the same as the ones obtained in the average game when the two games are not very different.
\end{abstract}

% insert suggested PACS numbers in braces on next line
% insert suggested keywords - APS authors don't need to do this

%\maketitle must follow title, authors, abstract, \pacs, and \keywords
\maketitle

\section{Introduction}
\label{Introduction}

How is it possible for cooperation to exist in a system dominated by competitive and self-interested individuals? This is one of the big open questions in science \cite{sciencespecial}. It becomes more interesting when we realize that cooperation is not exclusively a human phenomena, it exists between different animal species \cite{interspeciecoop}, within very organized insect societies and amongst kin \cite{traulsen2008groupsel}. If we generalize cooperation, using the framework of game theory, even between cells and RNA the concept can be used to model the emergence of organization in more complex and efficient forms \cite{rnaprisionerd,coopinevo}.

In  competitive systems it would be reasonable to think that selfish behaviour would be the best choice of strategy, so how can cooperation spontaneously emerge? Evolutionary Game Theory is often used to analyze such questions, specifically using the so called Dilemma Games \cite{szabo2007review,nowakpd}. In such games players can cooperate to obtain a payoff, but there's a temptation to betray the cooperator and get a better payoff.

One of the simplest cases in game theory is the two-player game with only two possible strategies. Each player can choose either to cooperate ($C$) or to Defect ($D$) in each round. Although one player can interact with many other players, each round is played only by two individuals. The payoff matrix ($\textbf{G}=(T,~R,~P,~S)$) represents how much the players gain from a single game and is normally state as: if both cooperated they receive $R$ (Reward), if both defected they receive $P$ (Punishment) and if they chose different strategies the defector receives $T$ (Temptation) and the cooperator receives $S$ (Sucker). These four parameters can describe different games  \cite{szabo2007review,nowak2006book,wardil2013mix}, being the Prisoners Dilemma (PD) the most canonical case, where $T>R>P>S$. This means that the best payoff is to defect when someone tries to cooperate with you. But if the whole population starts to defect the total payoff is lower than of a cooperating population. The Snow-Drift game (SD) happens when $T>R>S>P$. Observe that in this case it is worst when both players defect. This is a common game in animal contests where the damage of escalating conflicts is usually higher than being exploited \cite{snowdriftbook}. The Stag-Hunt (SH) happens when $R>T>P>S$. Now it is better to imitate your opponent. This game can describe the behaviour of cooperative associations, where one can exploit the other, but the best results are achieved by cooperation \cite{staghuntbook}. The last case is the Harmony-Game (HG), characterized by $R>T$ and $S>P$. In this case  it is always best to cooperate.

The evolution of cooperation in well-mixed populations can be described by mean-field equations, like the replicator equation \cite{szabo2007review}. It predicts that cooperation cannot survive, a conclusion that does not seems to happen in nature \cite{nowak2006book,szabo2007review,axelrodbook}. Many mechanisms have been proposed to explain the survival of cooperation: direct and indirect reciprocity \cite{axelrodbook,Sigmund2012indirect,epljaff,prelucasjaff}, volunteering \cite{hauert2002volunter}, spatial selection \cite{nowak1992spatial}, multi-level and kin selection \cite{traulsen2008groupsel}. The essential feature of these mechanisms is that cooperators interact more often between themselves than with defectors.  For example, spatial reciprocity happens when individuals are located in the vertices of a graph and can interact only with their neighbours. Cooperators spontaneously form clusters of cooperation that survive in a sea of defectors for a wide range of parameters \cite{szabo2007review}. Most models assume that only one game is repeatedly played in all interactions. However, there is nothing a priori precluding the coexistence of different games in a population.  For example, players can have different perceptions of  the rewards, giving rise to a multi-game scenario. In this case, individuals play asymmetric games, where the payoff matrix of one player  is different from the payoff matrix of his co-player \cite{multigames,multigames2}. Here, we study a model where different payoff matrices are randomly assigned to each interaction  -- the \textit{mixed-game model} \cite{wardil2013mix}. Different from the multi-game model, here the game is symmetric, that is, in each  round both players use the same payoff matrix.

The mixed-game model is presented  in the next section. In section III we  develop  the master equation approach for mixed-games. We show that these equations are identical to the ones where the evolution is driven by a single game with payoff given by the average of the two games (average game), when particular conditions hold. In section IV we use Monte-Carlo simulations to investigate the evolution of the cooperators for mixed-games in a two-dimensional lattice.  We also analyze how robust is the result of one cooperative game if there is a chance of a more selfish game be played and compare the mixed-game model to the average game. The relation of the master equation with the replicator equation is discussed in Appendix A. Appendix B contains the development of the pair-approximation model used to study the mixed-games. Finally, we present our conclusions in the last section.

\section{The  model}
\label{Model}

\begin{figure}[!h]
\includegraphics[scale=0.35]{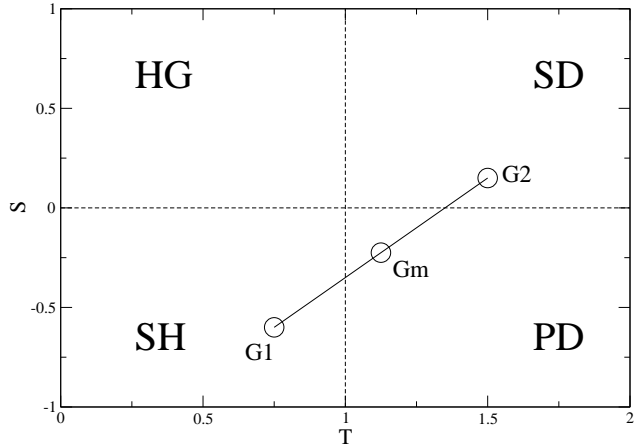}
\caption{Parameter space in T and S for the possible payoff matrices. Using R=1 and P=0 we have the four games (HG, SD, PD, SH) well defined in each quadrant. For every two single games (SH=$\textbf{G}_{1}$, and  SD=$\textbf{G}_{2}$, for example) there will be a mean game that is in the medium point between the two games. This mean game can be a totally different game (PD), as is shown in the example above.}
\label{fig_parameterspace} 
\end{figure}

We simulate the evolution of cooperation in populations structured on two-dimensional square lattices with first neighbour interactions and periodic boundary conditions. Initially, each individual is assigned to one strategy ($C$ or $D$) with equal probability. The strategies determine the payoff obtained on pairwise interactions. We choose to parametrize R=1, P=0, $T\in[0,2]$ and $S\in[-1,1]$ to obtain an useful parameter space that contains all four games (HG, SH, PD and SD), as can be seeing in figure \ref{fig_parameterspace}. By doing so we can analyze how the fraction of cooperators in the stationary state ($\rho(t\rightarrow\infty)$) is affected by such parameters and if there's is any phase transition.

The evolutionary process of mixed-games is composed of two parts: the game phase and the imitation phase. In the \textit{game phase}, each individual  plays one round of a game with each one of  his four neighbours and the payoff obtained in each interaction is added to the cumulative payoff of each player. The games to be played  are randomly assigned to each interaction: $\textbf{G}_{1}=(T_1,S_1)$ with probability $w$ or $\textbf{G}_{2}=(T_2,S_2)$ with probability $1-w$. In the \textit{imitation phase}, individuals may update their strategies by imitating the strategy adopted by more successful neighbours: the focal individual $i$ randomly chooses one neighbour $j$ to copy his strategy with probability $p(\Delta u_{ij})$ depending on the payoff difference $\Delta u_{ij}=u_j-u_i$, where $u_i$ is the cumulative payoff of player $i$.  For the strategy update probabilities, we use the Fermi-Dirac and the Proportional Imitation rule \cite{szabo2007review,hauertpair}:
\begin{equation}\label{fermi}
p(\Delta u_{ij})=1/ \left( 1+e^{-\beta(u_{j}-u_{i})/K} \right)
\end{equation}
\begin{equation}\label{prop}
p(\Delta u_{ij})=\rm{max}[(u_{j}-u_{i}),0]/b~~,
\end{equation}
respectively. The parameter $b$ is a normalizing factor and $K$ is the parameter controlling the strength of selection pressure on strategies \cite{nowak2006book}. 

The game and the imitation phases can be implemented  synchronously, partially asynchronously, or totally asynchronously \cite{nowak2006book,montecarlogames}. In the \textit{synchronous protocol}, in each Monte Carlo step (MCS) the cumulative payoffs are set to zero, games are randomly and independently  assigned to each one of the interactions, and the cumulative payoff of every player is calculated. Then all players simultaneously proceed to the imitation phase. In the \textit{totally asynchronous protocol} the cumulative payoffs are set to zero, a randomly chosen focal player and his four neighbours play randomly and independently assigned games with their respective neighbours, the cumulative payoff of these five players is calculated, and, finally only the focal player proceeds to the imitation phase. This process is repeated until all nodes have equal chance to update their strategies at least once -- this defines one Monte Carlo step (MCS). In the \textit{partially asynchronous protocol}, the games that each player will play with its neighbour are assigned only once at the beginning of each MCS -- as in the synchronous protocol -- and then strategies are update asynchronously. In this work, all protocols  yields the same qualitative results.

We measure the cooperation level as the average number of cooperators $\rho$ in the stationary state. We compare the evolution of cooperation in the mixed-game  case  to the case where only the average  game $\textbf{G}_m= \langle \textbf{G}\rangle =  w\textbf{G}_{1}+(1-w) \textbf{G}_{2}$ is played and to the cases where only $\textbf{G}_{1}$ or $\textbf{G}_{2}$ is played.

\section{Master equation approach}
Perhaps the first interesting question is about the equivalence between the mixed-games and the average game. In this section we investigate such equivalence for the asynchronous update. Obviously we are unable to solve the master equation exactly. However, we are able to establish conditions that allow us to show this equivalence. These condition are : (i) the strategies are statistically independent of the game distribution and (ii) the transition rates are linear functions of the payoffs. The main  result of this section can be used as a guideline for more general problems.

At each time $t$ such system is characterized by the strategy configuration $\{s\}=\{s_1,s_2,...s_N\}$, where $s_i$ is the strategy of player $i$, and by the game assignment configuration $\{g\}=\{g_{12},g_{13},\ldots\}$. The variable $g_{ij}$ is uniformly distributed over the discrete set of values $\{1,2\}$, determining which one of the two games, $\textbf{G}_{1}$ or $\textbf{G}_{2}$,  are to be played in the interaction between player $i$ and its next neighbour $j$. Note also that each link $i,j$ is statistically independent of the other links. Let $P(\{s\},\{g\},t)$ be the probability to find  the mixed-game system in the configuration $\{s\}$ and $\{g\}$ at time $t$. The time evolution of this system is given by
\begin{eqnarray}\nonumber
\label{mastereq}
\frac{d}{dt}P(\{s\},\{g\},t)=\sum_{\{s\}',\{g\}'} {P(\{s\}',\{g\}',t)}W(\{s\}',\{g\}'\rightarrow\{s\},\{g\}) \\ -{P(\{s\},\{g\},t)}W(\{s\},\{g\}\rightarrow\{s\}',\{g\}')~~,
\end{eqnarray}
where $W(\{s\}',\{g\}'\rightarrow\{s\},\{g\})$ is the transition rate from the state $(\{s\}',\{g\}')$ to $(\{s\},\{g\})$. If  the initial game assignment probability distribution is the stationary distribution $\Theta = \prod_{ij} \theta(g_{ij})$, $\Theta$ is independent of  time. Here we have that $\theta(g)=w\delta_{g,1}+(1-w)\delta_{g,2}$. Supposing the condition (i) that the strategy variables $\{s\}$ are statistically independent of the game variables $\{g\}$, we can write that $P(\{s\},\{g\},t)=P(\{s\},t)P(\{g\})$. To simplify the the notation, from now on we write $s$ and $g$ instead of $\{s\}$ and $\{g\}$. The master equation can now be written as
\begin{eqnarray}\nonumber
\label{mastereq2}
P(g)\frac{d}{dt}P(s,t)=\sum_{s',g'} {P(s',t)P(g')}W(s',g'\rightarrow s,g) \\ 
-{P(s,t)P(g)}W(s,g\rightarrow s',g')~~.
\end{eqnarray}
The mean value of a function $f$ averaged over all game configurations is $\left\langle f \right\rangle_g=\sum_{g}fP(g)$. Using this, we can sum in $g$ in both sides of equation (\ref{mastereq2}) to obtain that
\begin{eqnarray}\nonumber
\frac{d}{dt}P(s,t)&=&\sum_{s',g}P(s',t)   \left[ \sum_{g'}   P(g')W(s',g'\rightarrow s,g) \right]  \\  \nonumber               &-&\sum_{s',g'}P(s,t)   \left[  \sum_{g} P(g)W(s,g\rightarrow s',g') \right] ~~.
\end{eqnarray}
Since the quantities inside square brackets are the averaged values of $W$ over $g$($g'$)  we have that
\begin{eqnarray}\nonumber
\label{mastereq7}
\frac{d}{dt}P(s,t)=&\sum&_{s'}P(s',t)   \left\langle \sum_{g} W(s',g'\rightarrow s,g) \right\rangle_{g'}  \\                 -&\sum&_{s'}P(s,t)   \left\langle \sum_{g'} W(s,g\rightarrow s',g') \right\rangle_{g}~~.
\end{eqnarray}

The transition rates $W$ will be obtained from each microscopical rule (proportional imitation, Fermi-Dirac, copy the best, etc.), and will depend on the payoff differences($\Delta u_{ij}$) of neighbouring sites, topology of the lattice and the copy mechanism. Since the dynamical process in a time step, in our model, always involves only one site, for example site $i$, copying the strategy of a neighbour, say $j$, all the transition rates in which the new $ g'$ has a link that is not connected to site $i$ are null. Moreover all these rules are explicitly independent of the future game to be played, making the summation inside the averages lead to a finite constant $\kappa$ that will only depend on the geometry of the lattice and on some aspects of the microscopic model. It follows that
\begin{equation}
\left\langle \sum_{g} W(s',g'\rightarrow s,g) \right\rangle_{g'}=\kappa \left\langle W(g',s'\rightarrow s) \right\rangle_{g'}~~.
\end{equation}
Notice that for simplicity we hide $\Delta u_{ij}$ from the transition rates $W(s',g'\rightarrow s,g;\Delta u_{ij})$ variables, although it depends on it. In regular lattices $\kappa$ is the same constant for every site and can be absorbed in the time variable and we get

\begin{equation}
\label{jaff1}
\frac{d}{dt}P(s,t)=\sum_{s'}P(s',t)   \left\langle W(g,s'\rightarrow s) \right\rangle_{g}   -\sum_{s'} P(s,t)  \left\langle W(g,s\rightarrow s') \right\rangle_{g}~~.
\end{equation}
If we use condition (ii), that the update rules are linear on the payoff, the average over the transition rules will become simply the the transition rates of the average payoff, namely
\begin{equation}
 \left\langle W(s',g'\rightarrow s,g;\Delta u_{ij})\right\rangle_{g}= W(s',g'\rightarrow s,g;\left\langle\Delta u_{ij}\right\rangle_{g})~~.
\end{equation}
 Using this relation in equation \ref{jaff1}, we obtain that 
\begin{equation}
\label{mastereq3}
\frac{d}{dt}P(s,t)=\sum_{s'}P(s',t) W(s'\rightarrow s;\langle {\textbf G}\rangle)   -\sum_{s'} P(s,t)   W(s\rightarrow s';\langle {\textbf G}\rangle) ~~,
\end{equation}
where $\langle {\textbf G}\rangle$ stands for the average payoff matrix. This equation implies that the time evolution of the configuration probability in a mixed-game with asynchronous update  is equal to the time evolution of a single game \cite{szabo2007review} with the transition rates evaluated by using the average payoff matrix. Note this equation also describes the classical evolution of a single game if we
use the appropriate transition rates.

It is worth mentioning that this result also holds for  more general game assignment distributions. Supposing that $\theta_{g}=\sum_{k=1}^n w_k\delta_{g,k}$, where $w_k$ is the probability of play the game ${\bf G}_k$ ($k=1,\dots ,n$), we can easily see that the argument remains the same. The only difference is that the average payoff matrix is now given by $\langle {\textbf G}\rangle =\sum_{k=1}^n w_k {\textbf G}_k$.

We emphasize that equation (\ref{mastereq3}) was obtained by assuming that the variables $\{s\}$ and $\{ g\}$ are not correlated, condition (i). Also the equivalence between the average and the mixed-games only holds for the master equation as long as the transitions rates are linear on the payoff, condition (ii).

\begin{figure}
\includegraphics[scale=0.48]{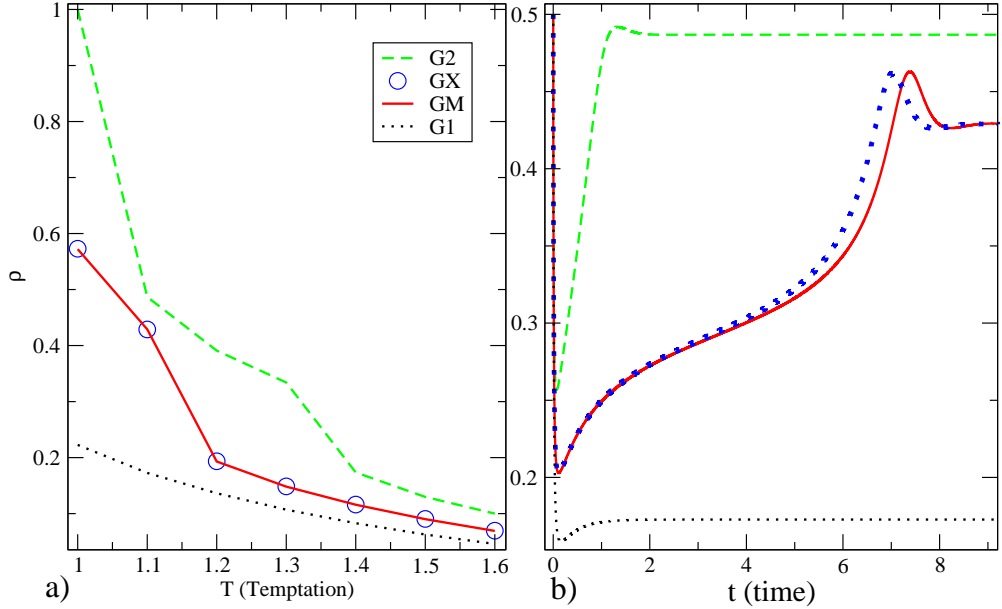}
\caption{(Color Online) Integration of the mean-field, pair-approximation, ODE by a Runge-Kutta algorithm. (a)  Phase transition of the stationary state value of cooperators in different single games ($G_1$, $G_2$ and $G_m$) and the mixed-game ($G_x$). $\textbf{G}_{m}$($S_{m}=0$) is the mean game of $\textbf{G}_{1}$($S_{1}=-0.1$) and $\textbf{G}_{2}$($S_{1}=0.1$), GX is the model where both $\textbf{G}_{1}$ and $\textbf{G}_{2}$ are played with equal probability. (b) Time evolution of cooperators for each game, here $T=1.1$ . Although the mixed-game behaves differently for small times, the fraction of cooperators in the stationary state is similar.}
\label{fig_meanfield} 
\end{figure}

In Appendix A we expand the replicator equation model to include mixed-games. The numerical solution of the master equation can be refined into more accurate models that consider spacial interactions. Appendix B shows the classic pair approximation \cite{matsudapair,hauertpair} and how we expanded it to include mixed-games in a square lattice. Our numerical solution shows that the mixed-game is equivalent to the mean game. But in  the pair approximation this only holds as long as the parameters (T and S) of the mixed-games are close enough from the mean game. The obtained results are shown in figure \ref{fig_meanfield}. 

\section{Monte Carlo simulations}

Let us now study the mixed-game model with populations structured in a two-dimensional square lattice. Since the variables $ g_{ij}$ defining the game between sites $i$ and $j$ obeys the stationary probability  distribution $\Theta= \prod_{ij} \{w\delta_{g_{ij},1}+(1-w)\delta_{g_{ij},2}\}$, they are independent of the strategy's variables. On the other hand, the probability of $s_i$ and $s_j$ have  particular values, $(C,C)$ for example, depends if the game between them is $G_1$ or $G_2$. Therefore condition (i) is not obeyed. Moreover, the transitions rates are not linear when we use the payoffs defined by equations (\ref{fermi}) and (\ref{prop}). This implies that condition (ii) is also not obeyed. To investigate the effect of breaking conditions (i) and (ii), we performed Monte-Carlo simulations on the square lattice. Simulations were performed on lattices with sizes(L) ranging from $100\times100$  to $500\times500$. All quantities are averaged over $20$ initial conditions and let termalize for at least 3000 Monte-Carlo steps for each run. 
\begin{figure}[h]
\includegraphics[scale=0.35]{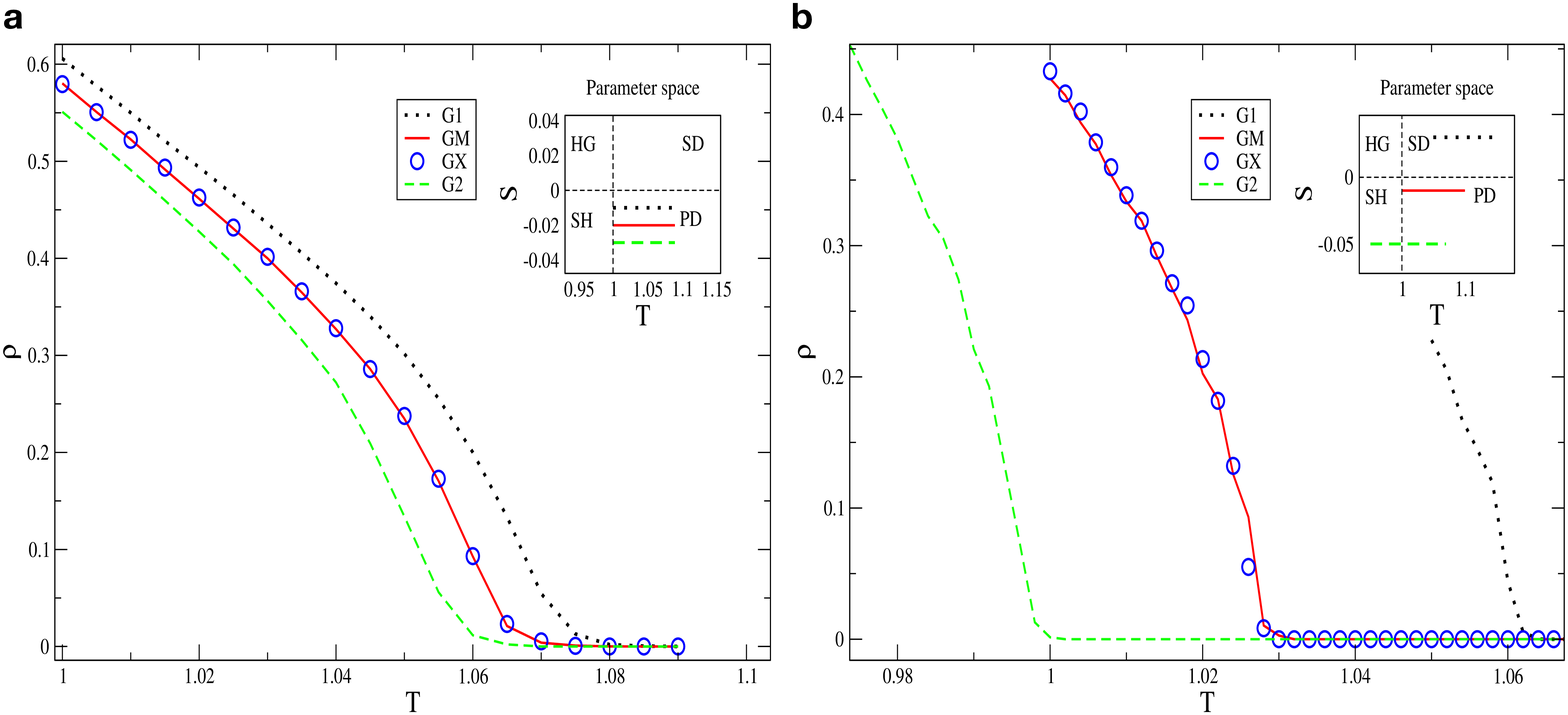}
\caption{(Color Online) (a) Phase transition of cooperators in different PD games for an asynchronous model,$L=100$. $\textbf{G}_{m}(S=-0.02)$ is the mean game of $\textbf{G}_{1}(S=-0.01)$ and $\textbf{G}_{2}(S=-0.03)$, GX is the model where both $\textbf{G}_{1}$ and $\textbf{G}_{2}$ are played. (b) Phase transition in different games for a synchronous model, $L=200$. Here $\textbf{G}_{1}(T>1.05,S=0.02)$ is a SD game, $\textbf{G}_{2}(T>0.95,S=-0.05)$ a SH and the mean game $\textbf{G}_{m}(T>1,S=-0.01)$ is a PD. The mix of those two different games still behaves as a PD.}
\label{fig2} 
\end{figure}
It turns out that if the difference between $\textbf{G}_{1}$ and $\textbf{G}_{2}$ is small, the evolution of cooperation in a mixed-game is equivalent to the average game  $\textbf{G}_{m}$, as shown in figure \ref{fig2}. Interestingly, figure \ref{fig2}-(b) shows that cooperation may thrive in the presence of disadvantageous payoff matrix values as long as the average game is a favourable one. Notice that the mixed-game behaves as the mean game even when the two games are of different kinds, as shown in figure \ref{fig2}-(b), where $\textbf{G}_{1}$ is a SD and $\textbf{G}_{2}$ is SH game.

\begin{figure}
\includegraphics[scale=0.36]{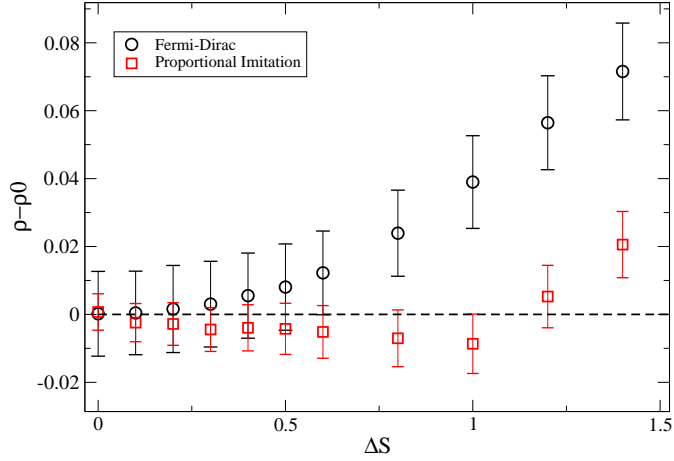}
\caption{(Color Online) Difference in the cooperation from the mean game and mixed-games versus the difference from the parameter $S_1$ and $S_2$. The dashed line corresponds to the mean game value of cooperation, squares point the data from the Proportional Imitation model and circles the Fermi-Dirac. Until $\Delta =1$ the Proportional Imitation model just fluctuates around the mean game value, while the Fermi-Dirac progressive rises. The cooperator differences are of $10^{-2}$ order.}
\label{fig_delta} 
\end{figure}

Figure \ref{fig_delta} shows the difference between  the fraction of cooperators for  the mixture of games $\textbf{G}_{1}=(1.04,S)$ and $\textbf{G}_{2}=(1.04,-S)$ and the fraction of cooperation for the corresponding average game  $\textbf{G}_{m}$. If the difference between the games $\textbf{G}_{1}$ and $\textbf{G}_{2}$  grows, then the equivalence no longer holds. Hence, our simulations suggest that, even if the update rules are non-linear, the equivalence between average and mixed-game may hold as long as the difference between the games $\textbf{G}_{1}$ and $\textbf{G}_{2}$ is small.
Nevertheless, the fraction of cooperation differs at most $0.1$. The small deviations contrast with the results obtained in \cite{multigames,multigames2}, where the authors  modelled multi-games as a result of different perceptions of the interactions. In their paper the games are asymmetric, meaning that the interface between cooperators and defectors can be beneficial for cooperators, strengthening even more the expansion of cooperation clusters. In our model, the symmetric games are randomly assigned every time step and the two players always have the same perception of the game. Hence, the cooperators on the interface can thrive only if the average game is cooperative.

\section{Conclusion}
We investigated under which circumstances the evolution of cooperation in the mixed-game model is equivalent to play a single game defined as the average between the two games in the mixed-game model. Using a master equation approach we showed that the time evolution of a mixed-game with asynchronous update  is equivalent to the average game if the transition rates are linear in the payoff and if there are no correlations between the game played and the state of the player. Using both synchronous and asynchronous Monte-Carlo simulations in the square lattice, we investigated the effects of (i) non-linear update rules and of (ii) correlations between the strategy and game assignment variables. We found that, as long as the differences between the two games are small, the mixed-game and the average game are equivalent. We showed also that assigning different games to interactions is not the same as to say that individuals have different asymmetric perceptions of the interaction.  

\label{Conclusion}

\begin{acknowledgements}
The authors  thank to CNPq and FAPEMIG, Brazilian agencies.
\end{acknowledgements}

%\newpage

\section{Appendix A}
\label{appendixA}	

We now derive the replicator equation for mixed-games. Suppose an infinite population of strategists where everybody interacts with everybody. Associating the mean payoff of a strategy with its capability of "reproduction" we can obtain mean field equations for the dynamics of the strategy population. The most simple cases comes without considering spatial effects and using just a single game being played. The replicator equation for this model can be written \cite{szabo2007review,repdinamics} in terms of games as
\begin{myequation}
\label{replieq}
\dot{x_{i}}=x_{i}[f_{i}(\vec{x})-\phi(\vec{x})]~~.
\end{myequation}
Here $x_i$ is the fraction of players using strategy $i$ (in our case $C$ or $D$), $f_{i}(\vec{x})$ can be regarded as the mean payoff of strategy $i$, and $\phi(\vec{x})$ is the total population average payoff. This mean field dynamics can be mapped into a master equation of a Markovian chain \cite{szabo2007review} that represents each player changing its state using some transition rule. It follows that
\begin{myequation}
\label{markoveq}
 \dot{x_{i}}=\sum_{j}[x_{j}\omega_{ji}(\vec{x})-x_{i}\omega_{ij}(\vec{x})]~~.
\end{myequation}
Here $\omega_{ij}(\vec{x})$ is the individual transition rate of a player from state $i$ to $j$, and can be obtained from the microscopic rule that states how one site copies the strategy of another. Many different microscopical rules have been proposed, such as Imitate the Best, Best response, Fermi-Dirac probability and Proportional Imitation rule. As can be expected, each one of this rules changes the results of the replicator dynamics, leading to different final results. The most common cases are the Maynard-Smith and Taylor-Johnsen equations \cite{maynardsmith,taylorjonker} which can be obtained from the Proportional Imitation and Moran process microscopical rules \cite{szabo2007review}.

Now we generalize this result to mixed-games. Being able to play two different games at each interaction, the payoff matrix $\textbf{G}$, will be now a statistical quantity, with mean value $\textbf{G}_{m}$. This leads to a change in the transition rates of the model, where the microscopic rule now depends on a probability distribution ($\Theta$) of games being played. Although we are changing the game being played, the replicator equation still maintain their dependence with the payoff (i.e. $f_{i}(\vec{x})$). If we average the replicator equation in the game assignment variables, we obtain that
\begin{myequation}
\label{average}
\left\langle\dot{x_{i}}\right\rangle_{g}=\left\langle\sum_{j}[x_{j} \omega_{ji}(\vec{x})-x_{i}\omega_{ij}(\vec{x})]\right\rangle_{g}=\sum_{j}[x_{j}\left\langle\omega_{ji}(\vec{x})\right\rangle_{g}-x_{i}\left\langle\omega_{ij}(\vec{x})\right\rangle_{g}]~~.
\end{myequation}
If we use any transition rule that is linear in $\textbf{G}$, this will lead  $\langle\omega_{ij}(\vec{x})\rangle = \omega_{ij}(\vec{x};\langle \textbf{G}\rangle)$ that is equivalent to playing the mean game $\textbf{G}_{m}=w\textbf{G}_{1}+(1-w)\textbf{G}_{2}$. 
This description  agree with the one obtained from  equation (\ref{mastereq3}). Note that these results are also valid if we use the more general game distribution
$\theta_{g}=\sum_{k=1}^n w_k\delta_{g,k}$.

\section{Appendix B}
\label{appendixB}	

The simplest way to describe a repeated game is by using a mean-field approximation. Here it is supposed that all individuals play with all other players in an infinite population. By doing this one can expect to obtain the mean payoff of a cooperator ($u_{c}$) or defector ($u_{d}$) using the payoff matrix and the abundance of other defectors or cooperators. In this simple model we have:
\begin{subequations}
\begin{align}
u_{c}&=R\rho_{c}+S\rho_{d} \nonumber \tag{B1}\\ 
u_{d}&=T\rho_{c}+P\rho_{d}~~, \tag{B2}
\end{align}
\end{subequations}
where $\rho_{c}$ and $\rho_{d}$ are the fraction of cooperators and defectors of the population. Inserting these payoffs in the replicator equation (\ref{replieq}) as the fitness of each strategy, we obtain the time evolution of such strategies as predicted by a mean-field approximation. This framework does not account for spacial effects such as lattices topologies or number of neighbours and results in a scenario where only one specie survives, depending on the parameters. There is no coexistence. Notice that in this model the payoff is a linear function of the parameters, so it's equivalent to play a mean game $\textbf{G}_{m}$ and averaging the payoffs of two mixed-games $\textbf{G}_{1}$ and $\textbf{G}_{2}$.

\begin{figure}
\includegraphics[scale=0.50]{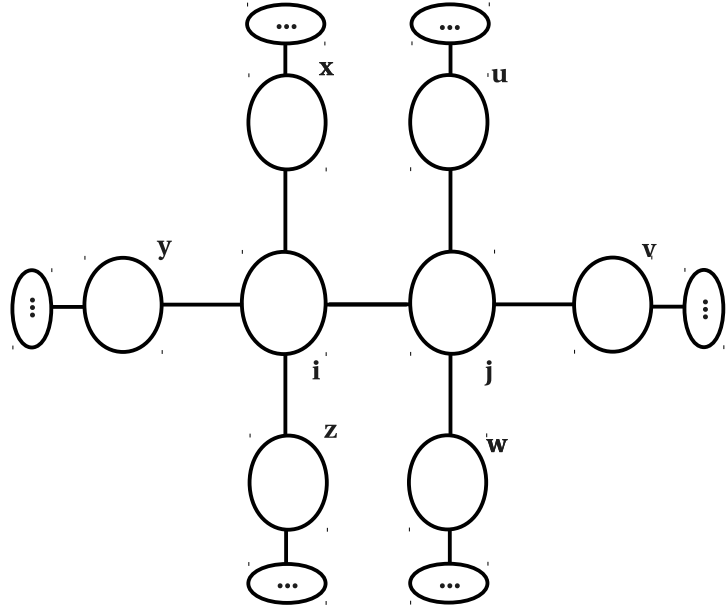}
\caption{The cluster structure used in the mean field-calculations. Although all seven main connections can give exact payoff between the sites, the second order neighbours (shown here as ``...'') payoffs needs to be calculated as a mean value using one site approximation. All eight sites can be either C or D and each connection can be in the states $\Gamma_{cc},\Gamma_{cd}$ and $\Gamma_{dd}$.}
\label{fig_pair} 
\end{figure}

The simplest way to take into account the lattice is by using the pair-approximation \cite{hauertpair,szabo2007review,matsudapair,pairbaalen}. Here we no longer suppose that all individuals are connected to everyone else. Instead we use a square lattice with interaction between the first next neighbours and analyze a cluster of two focal sites ($i$ and $j$) and their nearest next neighbours ($x,y,z$ and $u,v,w$) as it is shown in figure \ref{fig_pair}. Sites i and j have their payoff calculated exactly as they play games with cluster sites, while the neighbouring sites have an exact payoff calculated with i or j and the other 3 games (the second neighbours sites) payoffs are calculated by a mean value using a one site approximation. 

\begin{figure}
\includegraphics[scale=.75]{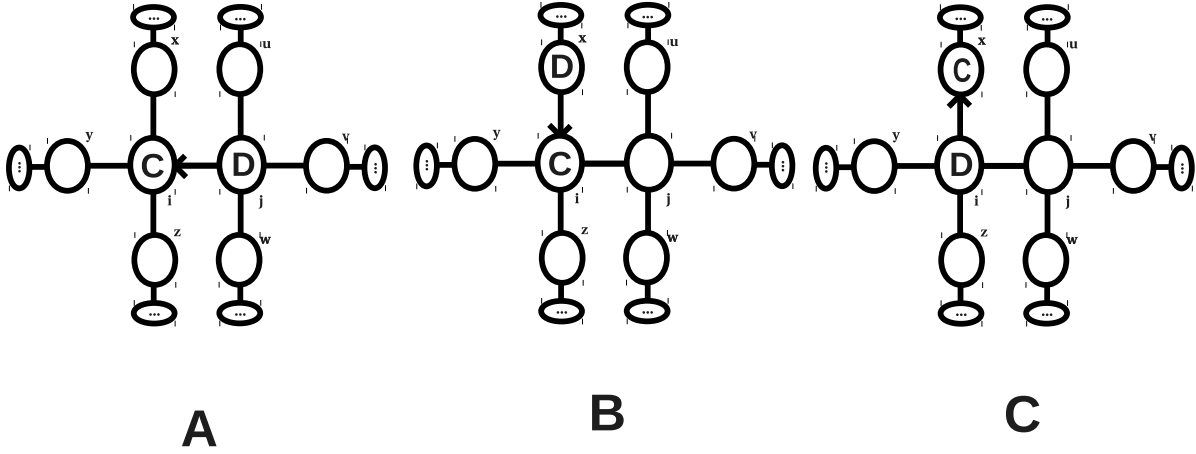}
\caption{Example of the three process that can change any link $\Gamma_{cd}$ to $\Gamma_{dd}$ in the cluster. Notice that this figure is just an example using process for the rate $W_{cd \rightarrow dd}$. Each of the other rates $W_{\alpha\beta \rightarrow \gamma\beta}$ have three similar process, exchanging $\alpha,\beta$ and $\gamma$ for the appropriate C or D.}
\label{fig_process} 
\end{figure}

To obtain the dynamics for this system we use the master equation (\ref{mastereq}). Now, instead of using $P(\{s\},\{g\},t)$ as the main variable, we use the probability that a connection between two sites is in the current state $\Gamma_{\alpha,\beta}$ ($\alpha$ and $\beta$ can be C or D in this case).  $W(\{s\},\{g\}\rightarrow\{s\}',\{g\}')$ become the rates that each link changes from one state to another ($cd \rightarrow cc$ for example). And we further simplify for the single game case ($\{g\}=\{g\}'$). Later we analyze the mixed-game models. This gives us three coupled ODE's that states how the fraction of possible connections in the cluster evolves:
\begin{subequations}
\begin{align}
\label{ode}
\dot{\Gamma}_{cc}&=\sum_{\{k\}} P(\{k\};cd)W_{cd \rightarrow cc}\{k\}-P(\{k\};cc)W_{cc\rightarrow cd}\{k\}~~, \tag{B3} \\ \nonumber
\dot{\Gamma}_{cd}&=\sum_{\{k\}} P(\{k\};cc)W_{cc \rightarrow cd}\{k\}+P(\{k\};dd)W_{dd \rightarrow cd}\{k\} \\ &~-P(\{k\};cd)W_{cd\rightarrow cc}\{k\}-P(\{k\};cd)W_{cd\rightarrow dd}\{k\}~~,\tag{B4} \\
\dot{\Gamma}_{dd}&=\sum_{\{k\}} P(\{k\};cd)W_{cd \rightarrow dd}\{k\}-P(\{k\};dd)W_{dd\rightarrow cd}\{k\}~~. \tag{B5}
\end{align}
\end{subequations}
Here $\left\{k \right\}$ is the configuration which describes the sites strategies, $P(\{k\};\alpha\beta)$ is the probability that the system is found in configuration $\{k\}$, given that exists at least one connection of the type $\alpha\beta$ and the summation occurs over every possible configuration that allows each one of the given transitions. The summation occurs over every possible cluster state, although in certain states the rate $W_{\alpha\beta \rightarrow \gamma\beta}$ that a pair can change is zero. We use the system constrains to reduce the problem to the resolution of only two ODE's, namely $\Gamma_{cc}+\Gamma_{cd}+\Gamma_{dd}=1$, $\rho_{c}=\Gamma_{cc}+\Gamma_{cd}/2$ and $\rho_{c}+\rho_{d}=1$. Each rate $W$ is calculated by counting every possible transition of the system from one state to the other and weighting it by the transition probability $p(u_{ij})$. Notice that as we look for changes in the links, not only i and j can change, but any of the seven links composing the cluster. In a broad sense each term in the ODE's is composed of only three terms:
\begin{equation}
P(\{k\};\alpha\beta)W_{\alpha\beta \rightarrow \gamma\beta}=A_{\alpha\beta}+B_{\alpha\beta}+C_{\alpha\beta}~~. \tag{B6}
\end{equation}
These terms represent the only possible processes that can change the links of the cluster in one interaction: A) the focal site copies the strategy of the other focal site, B) the focal site copies the strategies of a second order neighbour, C) the second order neighbour copies the strategy of focal site. This can be seen in figure \ref{fig_process}. These processes have a general form that depends on the two sites that will remain fixed in the summation (i.e. $\alpha,\beta$ being fixed C or D depending on each rate $W_{\alpha\beta \rightarrow \gamma\beta}$ )
\begin{subequations}
\begin{align}
\label{rates}
A_{(i=\alpha,j=\beta)} &= \sum_{x,y,z}\sum_{v,w,u} \eta_{\beta}\left[  \frac{\Gamma_{ix}\Gamma_{iy}\Gamma_{iz}}{\Gamma_{i}^{3}} \Gamma_{ij} \frac{\Gamma_{ju}\Gamma_{jv}\Gamma_{jw}}{\Gamma_{j}^{3}} \right] p(\Delta u_{ij})~~, \tag{B7} \\
B_{(i=\alpha,x=\beta)} &= 3\sum_{y,z}\sum_{v,w,u}\sum_{j} \eta_{\beta}\left[  \frac{\Gamma_{ix}\Gamma_{iy}\Gamma_{iz}}{\Gamma_{i}^{3}} \Gamma_{ij} \frac{\Gamma_{ju}\Gamma_{jv}\Gamma_{jw}}{\Gamma_{j}^{3}} \right] p(\Delta u_{ix})~~, \tag{B8} \\
C_{(i=\beta,x=\alpha)} &= 3\sum_{y,z}\sum_{v,w,u}\sum_{j} \left[  \frac{\Gamma_{ix}\Gamma_{iy}\Gamma_{iz}}{\Gamma_{i}^{3}} \Gamma_{ij} \frac{\Gamma_{ju}\Gamma_{jv}\Gamma_{jw}}{\Gamma_{j}^{3}} \right] p(\Delta u_{xi})~~. \tag{B9}
\end{align}
\end{subequations}
Here we multiplied by 3 the terms B and C to account for the number of equivalent repetitions of the same configuration that can occur due to symmetries.  $\eta_{\beta}$ is the number of neighbours of the current site in the same state as the site being copied (i.e. the number of links that will change if the current site changes its configuration). $\eta_{\beta}$ is always one for the C process, as it always changes just one link in the cluster. The quantity inside the square brackets gives the probability that the entire cluster will be in a given configuration $\left\{k \right\}=(x,y,z,v,w,u,i,j)$. Lastly $p(\Delta u_{ab})$ is given by the chosen microscopic transition rule, and gives the probability (given such configuration of the cluster) that the selected site ($a$) will copy its neighbour's ($b$) strategy, based on their payoff differences $\Delta u_{ab}$.
The payoff of any focal site is easy to be exactly obtained as $u_i=G(i,x)+G(i,y)+G(i,z)+G(i,j)$ (\textbf{G} is the game matrix). For the payoff of any neighbouring site x,y or z we need to get the mean payoffs using a one site approximation, namely
\begin{subequations}
\begin{align}
\bar{u}_{c}&=S+\rho_d^3(3S)+3\rho_d^2\rho_c(2S+R)+3\rho_d\rho_c^2(S+2R)+\rho_c^3(3R)~~, \tag{B10}\\
\bar{u}_{d}&=T+\rho_d^3(3\rho)+3\rho_d^2\rho_c(2\rho+T)+3\rho_d\rho_c^2(\rho+2T)+\rho_c^3(3T)~~.\tag{B11}
\end{align}
\end{subequations}
Here $\bar{u}_{c}$ is the mean payoff when the neighbour site is a cooperator and the focal site is a defector, and $\bar{u}_{d}$ is the payoff of the opposite situation.

The counting of every possibility of a site changing its strategy is complicated and give rise to about $2^{8}$ separated terms. Nonetheless, after determining a closed form for $A$, $B$ and $C$ for every rate $W$, it's a simple computational task to obtain all terms of $W_{\alpha\beta \rightarrow \gamma\beta}$. By doing so we obtain two coupled ODE's that can be numerically solved. We emphasize that this kind of mean-field results were already obtained in the literature \cite{hauertpair,szabo2007review,matsudapair,pairbaalen}.

Returning to the mixed-game problem, we used two possible games ($G_1$ and $G_2$) being played at each interaction. To insert this in the mean-field model with cluster approximation we have now to reconsider all the possible site payoffs. Instead of having just one payoff matrix being used for each interaction, we now have several different matrix weighted by their probability. This means that we have to sum over all the possible payoffs obtained with each site's neighbour in each game. It's equivalently, in our rate terms $W$, to change the payoff function $p(u_{\alpha \beta})$ into a sum over all possible games ($\sum_{\{g\}}f_g(u_{\alpha \beta})$) between the sites to be considered and its neighbours. It is important to notice however that the summation occurs over every neighbour of $\alpha$ and $\beta$ as the payoff of each site is determined by the sum of games played with every neighbour.

As each of the two sites being considered for each rate $W$ have 4 neighbours, this give rise to $2^6$ possibilities of different game configurations on the cluster. Notice however that the ODE's remains covariant as $\Gamma$ and the symmetry constants do not explicitly depend on the game being played. We can average over $\{g\}$, a vector that gives the configuration of each game being played, using it's probability distribution $\Theta(g)$. By doing so we obtain the new rates
\begin{equation}
\label{meanmix}
W_{\alpha\beta \rightarrow \gamma\beta}\left(\{k\};f\right)  \Rightarrow   \sum_{\{g\}}\Theta(g) W_{\alpha\beta \rightarrow \gamma\beta}'\left(\{k\};f_g\right)~~. \tag{B12}
\end{equation}
Making this change in the ODE and using a fourth order Runge-Kutta integration, we obtained the numerical results displayed in figure \ref{fig_meanfield}. Notice that as the payoff of each site depends on four other neighbours we need to consider every configuration of games being played by the sites in each term of $W$. This is why we cannot just sum up linearly the two different game matrices to obtain a mean payoff.

% Create the reference section using BibTeX:
\bibliography{myrefs}

%merlin.mbs apsrev4-1.bst 2010-07-25 4.21a (PWD, AO, DPC) hacked
%Control: key (0)
%Control: author (8) initials jnrlst
%Control: editor formatted (1) identically to author
%Control: production of article title (-1) disabled
%Control: page (0) single
%Control: year (1) truncated
%Control: production of eprint (0) enabled
\begin{thebibliography}{26}%
\makeatletter
\providecommand \@ifxundefined [1]{%
 \@ifx{#1\undefined}
}%
\providecommand \@ifnum [1]{%
 \ifnum #1\expandafter \@firstoftwo
 \else \expandafter \@secondoftwo
 \fi
}%
\providecommand \@ifx [1]{%
 \ifx #1\expandafter \@firstoftwo
 \else \expandafter \@secondoftwo
 \fi
}%
\providecommand \natexlab [1]{#1}%
\providecommand \enquote  [1]{``#1''}%
\providecommand \bibnamefont  [1]{#1}%
\providecommand \bibfnamefont [1]{#1}%
\providecommand \citenamefont [1]{#1}%
\providecommand \href@noop [0]{\@secondoftwo}%
\providecommand \href [0]{\begingroup \@sanitize@url \@href}%
\providecommand \@href[1]{\@@startlink{#1}\@@href}%
\providecommand \@@href[1]{\endgroup#1\@@endlink}%
\providecommand \@sanitize@url [0]{\catcode `\\12\catcode `\$12\catcode
  `\&12\catcode `\#12\catcode `\^12\catcode `\_12\catcode `\%12\relax}%
\providecommand \@@startlink[1]{}%
\providecommand \@@endlink[0]{}%
\providecommand \url  [0]{\begingroup\@sanitize@url \@url }%
\providecommand \@url [1]{\endgroup\@href {#1}{\urlprefix }}%
\providecommand \urlprefix  [0]{URL }%
\providecommand \Eprint [0]{\href }%
\providecommand \doibase [0]{http://dx.doi.org/}%
\providecommand \selectlanguage [0]{\@gobble}%
\providecommand \bibinfo  [0]{\@secondoftwo}%
\providecommand \bibfield  [0]{\@secondoftwo}%
\providecommand \translation [1]{[#1]}%
\providecommand \BibitemOpen [0]{}%
\providecommand \bibitemStop [0]{}%
\providecommand \bibitemNoStop [0]{.\EOS\space}%
\providecommand \EOS [0]{\spacefactor3000\relax}%
\providecommand \BibitemShut  [1]{\csname bibitem#1\endcsname}%
\let\auto@bib@innerbib\@empty
%</preamble>
\bibitem [{\citenamefont {E.}(2005)}]{sciencespecial}%
  \BibitemOpen
  \bibfield  {author} {\bibinfo {author} {\bibfnamefont {P.}~\bibnamefont
  {E.}},\ }\href@noop {} {\bibfield  {journal} {\bibinfo  {journal} {Science}\
  }\textbf {\bibinfo {volume} {309}},\ \bibinfo {pages} {93} (\bibinfo {year}
  {2005})}\BibitemShut {NoStop}%
\bibitem [{\citenamefont {Doebeli}\ and\ \citenamefont
  {Knowlton}(1998)}]{interspeciecoop}%
  \BibitemOpen
  \bibfield  {author} {\bibinfo {author} {\bibfnamefont {M.}~\bibnamefont
  {Doebeli}}\ and\ \bibinfo {author} {\bibfnamefont {N.}~\bibnamefont
  {Knowlton}},\ }\href@noop {} {\bibfield  {journal} {\bibinfo  {journal}
  {Proceedings of the National Academy of Sciences}\ }\textbf {\bibinfo
  {volume} {95}},\ \bibinfo {pages} {8676} (\bibinfo {year}
  {1998})}\BibitemShut {NoStop}%
\bibitem [{\citenamefont {Traulsen}\ \emph {et~al.}(2008)\citenamefont
  {Traulsen}, \citenamefont {Shoresh},\ and\ \citenamefont
  {Nowak}}]{traulsen2008groupsel}%
  \BibitemOpen
  \bibfield  {author} {\bibinfo {author} {\bibfnamefont {A.}~\bibnamefont
  {Traulsen}}, \bibinfo {author} {\bibfnamefont {N.}~\bibnamefont {Shoresh}}, \
  and\ \bibinfo {author} {\bibfnamefont {M.~A.}\ \bibnamefont {Nowak}},\
  }\href@noop {} {\bibfield  {journal} {\bibinfo  {journal} {Bulletin of
  mathematical biology}\ }\textbf {\bibinfo {volume} {70}},\ \bibinfo {pages}
  {1410} (\bibinfo {year} {2008})}\BibitemShut {NoStop}%
\bibitem [{\citenamefont {Turner}\ and\ \citenamefont
  {Chao}(1999)}]{rnaprisionerd}%
  \BibitemOpen
  \bibfield  {author} {\bibinfo {author} {\bibfnamefont {P.~E.}\ \bibnamefont
  {Turner}}\ and\ \bibinfo {author} {\bibfnamefont {L.}~\bibnamefont {Chao}},\
  }\href@noop {} {\bibfield  {journal} {\bibinfo  {journal} {Nature}\ }\textbf
  {\bibinfo {volume} {398}},\ \bibinfo {pages} {441} (\bibinfo {year}
  {1999})}\BibitemShut {NoStop}%
\bibitem [{\citenamefont {Szathm{\'a}ry}\ and\ \citenamefont
  {Smith}(1995)}]{coopinevo}%
  \BibitemOpen
  \bibfield  {author} {\bibinfo {author} {\bibfnamefont {E.}~\bibnamefont
  {Szathm{\'a}ry}}\ and\ \bibinfo {author} {\bibfnamefont {J.~M.}\ \bibnamefont
  {Smith}},\ }\href@noop {} {\bibfield  {journal} {\bibinfo  {journal}
  {Nature}\ }\textbf {\bibinfo {volume} {374}},\ \bibinfo {pages} {227}
  (\bibinfo {year} {1995})}\BibitemShut {NoStop}%
\bibitem [{\citenamefont {Szab{\'o}}\ and\ \citenamefont
  {Fath}(2007)}]{szabo2007review}%
  \BibitemOpen
  \bibfield  {author} {\bibinfo {author} {\bibfnamefont {G.}~\bibnamefont
  {Szab{\'o}}}\ and\ \bibinfo {author} {\bibfnamefont {G.}~\bibnamefont
  {Fath}},\ }\href@noop {} {\bibfield  {journal} {\bibinfo  {journal} {Physics
  Reports}\ }\textbf {\bibinfo {volume} {446}},\ \bibinfo {pages} {97}
  (\bibinfo {year} {2007})}\BibitemShut {NoStop}%
\bibitem [{\citenamefont {Nowak}\ and\ \citenamefont
  {Sigmund}(1990)}]{nowakpd}%
  \BibitemOpen
  \bibfield  {author} {\bibinfo {author} {\bibfnamefont {M.}~\bibnamefont
  {Nowak}}\ and\ \bibinfo {author} {\bibfnamefont {K.}~\bibnamefont
  {Sigmund}},\ }\href {\doibase 10.1007/BF00049570} {\bibfield  {journal}
  {\bibinfo  {journal} {Acta Applicandae Mathematica}\ }\textbf {\bibinfo
  {volume} {20}},\ \bibinfo {pages} {247} (\bibinfo {year} {1990})}\BibitemShut
  {NoStop}%
\bibitem [{\citenamefont {Nowak}(2006)}]{nowak2006book}%
  \BibitemOpen
  \bibfield  {author} {\bibinfo {author} {\bibfnamefont {M.~A.}\ \bibnamefont
  {Nowak}},\ }\href@noop {} {\emph {\bibinfo {title} {Evolutionary dynamics}}}\
  (\bibinfo  {publisher} {Harvard University Press},\ \bibinfo {year}
  {2006})\BibitemShut {NoStop}%
\bibitem [{\citenamefont {Wardil}\ and\ \citenamefont
  {da~Silva}(2013)}]{wardil2013mix}%
  \BibitemOpen
  \bibfield  {author} {\bibinfo {author} {\bibfnamefont {L.}~\bibnamefont
  {Wardil}}\ and\ \bibinfo {author} {\bibfnamefont {J.~K.}\ \bibnamefont
  {da~Silva}},\ }\href@noop {} {\bibfield  {journal} {\bibinfo  {journal}
  {Chaos, Solitons \& Fractals}\ }\textbf {\bibinfo {volume} {56}},\ \bibinfo
  {pages} {160} (\bibinfo {year} {2013})}\BibitemShut {NoStop}%
\bibitem [{\citenamefont {Rapoport}(1999)}]{snowdriftbook}%
  \BibitemOpen
  \bibfield  {author} {\bibinfo {author} {\bibfnamefont {A.}~\bibnamefont
  {Rapoport}},\ }\href@noop {} {\emph {\bibinfo {title} {Two-person game
  theory}}}\ (\bibinfo  {publisher} {Courier Dover Publications},\ \bibinfo
  {year} {1999})\BibitemShut {NoStop}%
\bibitem [{\citenamefont {Skyrms}(2004)}]{staghuntbook}%
  \BibitemOpen
  \bibfield  {author} {\bibinfo {author} {\bibfnamefont {B.}~\bibnamefont
  {Skyrms}},\ }\href@noop {} {\emph {\bibinfo {title} {The stag hunt and the
  evolution of social structure}}}\ (\bibinfo  {publisher} {Cambridge
  University Press},\ \bibinfo {year} {2004})\BibitemShut {NoStop}%
\bibitem [{\citenamefont {Axelrod}(1984)}]{axelrodbook}%
  \BibitemOpen
  \bibfield  {author} {\bibinfo {author} {\bibfnamefont {R.}~\bibnamefont
  {Axelrod}},\ }\href@noop {} {\emph {\bibinfo {title} {THE EVOLUTION 0F
  COOPERATION}}}\ (\bibinfo {year} {1984})\BibitemShut {NoStop}%
\bibitem [{\citenamefont {Sigmund}(2012)}]{Sigmund2012indirect}%
  \BibitemOpen
  \bibfield  {author} {\bibinfo {author} {\bibfnamefont {K.}~\bibnamefont
  {Sigmund}},\ }\href {\doibase http://dx.doi.org/10.1016/j.jtbi.2011.03.024}
  {\bibfield  {journal} {\bibinfo  {journal} {Journal of Theoretical Biology}\
  }\textbf {\bibinfo {volume} {299}},\ \bibinfo {pages} {25 } (\bibinfo {year}
  {2012})}\BibitemShut {NoStop}%
\bibitem [{\citenamefont {Wardil}\ and\ \citenamefont
  {da~Silva}(2009)}]{epljaff}%
  \BibitemOpen
  \bibfield  {author} {\bibinfo {author} {\bibfnamefont {L.}~\bibnamefont
  {Wardil}}\ and\ \bibinfo {author} {\bibfnamefont {J.~K.~L.}\ \bibnamefont
  {da~Silva}},\ }\href {http://stacks.iop.org/0295-5075/86/i=3/a=38001}
  {\bibfield  {journal} {\bibinfo  {journal} {EPL (Europhysics Letters)}\
  }\textbf {\bibinfo {volume} {86}},\ \bibinfo {pages} {38001} (\bibinfo {year}
  {2009})}\BibitemShut {NoStop}%
\bibitem [{\citenamefont {Wardil}\ and\ \citenamefont
  {da~Silva}(2010)}]{prelucasjaff}%
  \BibitemOpen
  \bibfield  {author} {\bibinfo {author} {\bibfnamefont {L.}~\bibnamefont
  {Wardil}}\ and\ \bibinfo {author} {\bibfnamefont {J.~K.~L.}\ \bibnamefont
  {da~Silva}},\ }\href {\doibase 10.1103/PhysRevE.81.036115} {\bibfield
  {journal} {\bibinfo  {journal} {Phys. Rev. E}\ }\textbf {\bibinfo {volume}
  {81}},\ \bibinfo {pages} {036115} (\bibinfo {year} {2010})}\BibitemShut
  {NoStop}%
\bibitem [{\citenamefont {Hauert}\ \emph {et~al.}(2002)\citenamefont {Hauert},
  \citenamefont {De~Monte}, \citenamefont {Hofbauer},\ and\ \citenamefont
  {Sigmund}}]{hauert2002volunter}%
  \BibitemOpen
  \bibfield  {author} {\bibinfo {author} {\bibfnamefont {C.}~\bibnamefont
  {Hauert}}, \bibinfo {author} {\bibfnamefont {S.}~\bibnamefont {De~Monte}},
  \bibinfo {author} {\bibfnamefont {J.}~\bibnamefont {Hofbauer}}, \ and\
  \bibinfo {author} {\bibfnamefont {K.}~\bibnamefont {Sigmund}},\ }\href@noop
  {} {\bibfield  {journal} {\bibinfo  {journal} {Science}\ }\textbf {\bibinfo
  {volume} {296}},\ \bibinfo {pages} {1129} (\bibinfo {year}
  {2002})}\BibitemShut {NoStop}%
\bibitem [{\citenamefont {Nowak}\ and\ \citenamefont
  {May}(1992)}]{nowak1992spatial}%
  \BibitemOpen
  \bibfield  {author} {\bibinfo {author} {\bibfnamefont {M.~A.}\ \bibnamefont
  {Nowak}}\ and\ \bibinfo {author} {\bibfnamefont {R.~M.}\ \bibnamefont
  {May}},\ }\href@noop {} {\bibfield  {journal} {\bibinfo  {journal} {Nature}\
  }\textbf {\bibinfo {volume} {359}},\ \bibinfo {pages} {29} (\bibinfo {year}
  {1992})}\BibitemShut {NoStop}%
\bibitem [{\citenamefont {Wang}\ \emph {et~al.}(2014)\citenamefont {Wang},
  \citenamefont {Szolnoki},\ and\ \citenamefont {Perc}}]{multigames}%
  \BibitemOpen
  \bibfield  {author} {\bibinfo {author} {\bibfnamefont {Z.}~\bibnamefont
  {Wang}}, \bibinfo {author} {\bibfnamefont {A.}~\bibnamefont {Szolnoki}}, \
  and\ \bibinfo {author} {\bibfnamefont {M.}~\bibnamefont {Perc}},\ }\href
  {\doibase 10.1103/PhysRevE.90.032813} {\bibfield  {journal} {\bibinfo
  {journal} {Phys. Rev. E}\ }\textbf {\bibinfo {volume} {90}},\ \bibinfo
  {pages} {032813} (\bibinfo {year} {2014})}\BibitemShut {NoStop}%
\bibitem [{\citenamefont {Szolnoki}\ and\ \citenamefont
  {Perc}(2014)}]{multigames2}%
  \BibitemOpen
  \bibfield  {author} {\bibinfo {author} {\bibfnamefont {A.}~\bibnamefont
  {Szolnoki}}\ and\ \bibinfo {author} {\bibfnamefont {M.}~\bibnamefont
  {Perc}},\ }\href {http://stacks.iop.org/0295-5075/108/i=2/a=28004} {\bibfield
   {journal} {\bibinfo  {journal} {EPL (Europhysics Letters)}\ }\textbf
  {\bibinfo {volume} {108}},\ \bibinfo {pages} {28004} (\bibinfo {year}
  {2014})}\BibitemShut {NoStop}%
\bibitem [{\citenamefont {Hauert}\ and\ \citenamefont
  {Szab{\'o}}(2005)}]{hauertpair}%
  \BibitemOpen
  \bibfield  {author} {\bibinfo {author} {\bibfnamefont {C.}~\bibnamefont
  {Hauert}}\ and\ \bibinfo {author} {\bibfnamefont {G.}~\bibnamefont
  {Szab{\'o}}},\ }\href@noop {} {\bibfield  {journal} {\bibinfo  {journal}
  {American Journal of Physics}\ }\textbf {\bibinfo {volume} {73}},\ \bibinfo
  {pages} {405} (\bibinfo {year} {2005})}\BibitemShut {NoStop}%
\bibitem [{\citenamefont {Huberman}\ and\ \citenamefont
  {Glance}(1993)}]{montecarlogames}%
  \BibitemOpen
  \bibfield  {author} {\bibinfo {author} {\bibfnamefont {B.~A.}\ \bibnamefont
  {Huberman}}\ and\ \bibinfo {author} {\bibfnamefont {N.~S.}\ \bibnamefont
  {Glance}},\ }\href@noop {} {\bibfield  {journal} {\bibinfo  {journal}
  {Proceedings of the National Academy of Sciences}\ }\textbf {\bibinfo
  {volume} {90}},\ \bibinfo {pages} {7716} (\bibinfo {year}
  {1993})}\BibitemShut {NoStop}%
\bibitem [{\citenamefont {Matsuda}\ \emph {et~al.}(1992)\citenamefont
  {Matsuda}, \citenamefont {Ogita}, \citenamefont {Sasaki},\ and\ \citenamefont
  {Sat{\=o}}}]{matsudapair}%
  \BibitemOpen
  \bibfield  {author} {\bibinfo {author} {\bibfnamefont {H.}~\bibnamefont
  {Matsuda}}, \bibinfo {author} {\bibfnamefont {N.}~\bibnamefont {Ogita}},
  \bibinfo {author} {\bibfnamefont {A.}~\bibnamefont {Sasaki}}, \ and\ \bibinfo
  {author} {\bibfnamefont {K.}~\bibnamefont {Sat{\=o}}},\ }\href@noop {}
  {\bibfield  {journal} {\bibinfo  {journal} {Progress of theoretical Physics}\
  }\textbf {\bibinfo {volume} {88}},\ \bibinfo {pages} {1035} (\bibinfo {year}
  {1992})}\BibitemShut {NoStop}%
\bibitem [{\citenamefont {Schuster}\ and\ \citenamefont
  {Sigmund}(1983)}]{repdinamics}%
  \BibitemOpen
  \bibfield  {author} {\bibinfo {author} {\bibfnamefont {P.}~\bibnamefont
  {Schuster}}\ and\ \bibinfo {author} {\bibfnamefont {K.}~\bibnamefont
  {Sigmund}},\ }\href {\doibase http://dx.doi.org/10.1016/0022-5193(83)90445-9}
  {\bibfield  {journal} {\bibinfo  {journal} {Journal of Theoretical Biology}\
  }\textbf {\bibinfo {volume} {100}},\ \bibinfo {pages} {533 } (\bibinfo {year}
  {1983})}\BibitemShut {NoStop}%
\bibitem [{\citenamefont {Smith}(1974)}]{maynardsmith}%
  \BibitemOpen
  \bibfield  {author} {\bibinfo {author} {\bibfnamefont {J.~M.}\ \bibnamefont
  {Smith}},\ }\href {\doibase http://dx.doi.org/10.1016/0022-5193(74)90110-6}
  {\bibfield  {journal} {\bibinfo  {journal} {Journal of Theoretical Biology}\
  }\textbf {\bibinfo {volume} {47}},\ \bibinfo {pages} {209 } (\bibinfo {year}
  {1974})}\BibitemShut {NoStop}%
\bibitem [{\citenamefont {Taylor}\ and\ \citenamefont
  {Jonker}(1978)}]{taylorjonker}%
  \BibitemOpen
  \bibfield  {author} {\bibinfo {author} {\bibfnamefont {P.~D.}\ \bibnamefont
  {Taylor}}\ and\ \bibinfo {author} {\bibfnamefont {L.~B.}\ \bibnamefont
  {Jonker}},\ }\href {\doibase http://dx.doi.org/10.1016/0025-5564(78)90077-9}
  {\bibfield  {journal} {\bibinfo  {journal} {Mathematical Biosciences}\
  }\textbf {\bibinfo {volume} {40}},\ \bibinfo {pages} {145 } (\bibinfo {year}
  {1978})}\BibitemShut {NoStop}%
\bibitem [{\citenamefont {Dieckmann}\ \emph {et~al.}(2000)\citenamefont
  {Dieckmann}, \citenamefont {Law},\ and\ \citenamefont {Metz}}]{pairbaalen}%
  \BibitemOpen
  \bibfield  {author} {\bibinfo {author} {\bibfnamefont {U.}~\bibnamefont
  {Dieckmann}}, \bibinfo {author} {\bibfnamefont {R.}~\bibnamefont {Law}}, \
  and\ \bibinfo {author} {\bibfnamefont {J.~A.}\ \bibnamefont {Metz}},\
  }\href@noop {} {\emph {\bibinfo {title} {The geometry of ecological
  interactions: simplifying spatial complexity}}}\ (\bibinfo  {publisher}
  {Cambridge University Press},\ \bibinfo {year} {2000})\ Chap.~\bibinfo
  {chapter} {19}, pp.\ \bibinfo {pages} {1--31}\BibitemShut {NoStop}%
\end{thebibliography}%

\end{document}